# Hurst Exponents, Markov Processes, and Fractional Brownian Motion

Joseph L. McCauley[+], Gemunu H. Gunaratne[++], and Kevin E. Bassler[+++]

Physics Department
University of Houston
Houston, Tx. 77204
jmccauley@uh.edu

[+]Senior Fellow
COBERA
Department of Economics
J.E.Cairnes Graduate School of Business and Public Policy
NUI Galway, Ireland

[++]Institute of Fundamental Studies
Kandy, Sri Lanka

[+++]Texas Center for Superconductivity
University of Houston
Houston, Texas

## Abstract

There is much confusion in the literature over Hurst exponents. Recently, we took a step in the direction of eliminating some of the confusion. One purpose of this paper is to illustrate the difference between fBm on the one hand and Gaussian Markov processes where $H \neq 1/2$ on the other. The difference lies in the increments, which are stationary and correlated in one case and nonstationary and uncorrelated in the other. The two- and one-point densities of fBm are constructed explicitly. The two-point density doesn't scale. The one-point density for a semi-infinite time interval is identical to that for a scaling Gaussian Markov process with $H \neq 1/2$ over a finite time interval. We conclude that both Hurst exponents and one point densities are inadequate for deducing the underlying dynamics from empirical data. We apply these conclusions in the end to make a focused statement about 'nonlinear diffusion'.

## 1. Introduction

The necessity of stationary increments for fBm has been emphasized in books [1] and papers [2] by mathematicians, but also is sometimes not stated [3]. Books [4] and papers [5] by physicists tend to ignore the question altogether and to assume, without justification and incorrectly, that $H \neq 1/2$ always implies long time correlations. In this paper we emphasize that the essential point in long time correlations is stationarity of the increments, not scaling. Scaling makes life simpler, when it occurs, but is irrelevant for correlations: classes of stochastic dynamical systems with long time correlations exist without scaling, while processes with no memory at all, Markov processes, can scale perfectly with a Hurst exponent $H \neq 1/2$ [6]. Our point is that evidence for scaling, taken alone, tells us nothing whatsoever about the existence of long time correlations. The basic question to be answered first, in both data analysis and theory, is: are the increments stationary or nonstationary? To achieve both unity and clarity, we begin with the mathematicians' usual definition of scaling of a stochastic process and then show how it leads naturally to the physicists' definition.

## 2. Hurst exponent scaling

We define scaling (self-similar processes) starting from the mathematicians' standpoint [1] and show that it's equivalent to our definition [6] in terms of densities.

A stochastic process x(t) is said to scale with Hurst exponent H if [1]

$$x(t) = t^H x(1), \quad (1)$$

where by equality we mean equality 'in distribution'. We next define what that means in practice.

The 1-point distribution P(x,t) reflects the statistics collected from many different runs of the time evolution of x(t) from a specified initial condition $x(t_o)$, but doesn't describe correlations or lack of same. The (one-point) density $f_1(x,t)$ of the distribution is defined by $f_1(x,t)=dP/dx$. Given any dynamical variable A(x,t), averages of A are calculated via

$$\langle A(t) \rangle = \int_{-\infty}^{\infty} A(x,t) f_1(x,t) dx. \quad (2)$$

We restrict to x-independent drift coefficients R(x,t)=R(t), $<x(t)>=x(t_o)$ where $t_o$ denotes the initial observation time, for a very good reason. Since the drift in the absolute average

$$\langle x(t) \rangle = x(t_o) + \int_{t_o}^{t} R(s) ds \quad (3)$$

depends only on t then it's trivial to remove it from the stochastic process by choosing instead of x(t) the Martingale variable z(t)=x(t)-∫R(s)ds. In what follows we will write 'x(t)' with the assumption that the drift has been removed, <x(t)>=x($t_o$). We will also take x($t_o$)=0, so that more generally our x(t) must be interpreted as x(t)-x($t_o$) if x($t_o$)≠0. I.e., we are generally using x(t) to mean the random variable z(t)=x(t)-x($t_o$)-∫Rdt.

From (1), the moments of x must obey

$$\langle x^n(t) \rangle = t^{nH} \langle x^n(1) \rangle = c_n t^{nH} \quad (4)$$

Combining this with

$$\langle x^n(t) \rangle = \int x^n f_1(x,t) dx \quad (5)$$

we obtain [6]

$$f_1(x,t) = t^{-H} F(u), \quad (6)$$

where the scaling variable is $u=x/t^H$. In particular, with no average drift, so that we can choose <x(t)>=x($t_o$)=0, the variance is simply

$$\sigma^2 = \langle x^2(t) \rangle = \langle x^2(1) \rangle t^{2H}. \quad (7)$$

This explains what is meant by Hurst exponent scaling, and also specifies what's meant that (1) holds 'in distribution'. In all that follows, equations in random variables x(t) like (1), solutions of stochastic differential equations, and increment equations as we shall write down in part 3 below, are all to be understood as holding 'in distribution'.

We ignore Levy distributions here because they are not indicated by our recent financial data analysis [7]. That analysis suggests diffusive processes, Markov processes. For discussions of Levy distributions, see Scalas et al [8,9].

Empirically, the best evidence for scaling would be a data collapse of the form $F(u)=t^H f_1(x,t)$. Next best but weaker is to look for scaling in a finite number of the moments $<x^n(t)>$. It's important to understand that Hurst exponent scaling, taken alone, tells us nothing about the underlying stochastic dynamics. In particular, scaling, taken alone, implies neither the presence nor absence of autocorrelations increments/displacements taken over nonoverlapping time intervals, as we will show next.

Selfsimilar processes (1) are strongly nonstationary: by (4), the moments do not approach constants, rather, the one-point density necessarily spreads in width forever, reflecting the continual loss of information about where x(t) lies [6]. But a nonstationary process may have either stationary or nonstationary *increments*.

## 3. Stationary vs. nonstationary increments

Stationary processes are often confused with stationary increments in the literature (see [6] for a discussion). Stationary increments are assumed without prior justification in data analyses. We define stationary and nonstationary increments and exhibit their implications for the question of long time autocorrelations, or complete lack of autocorrelations. We show that the question of stationary increments, not scaling, is central for the existence of long time correlations.

Stationary increments $\Delta x(t,T)$ of a nonstationary process $x(t)$ are defined by

$$x(t + T) - x(t) = x(T), \qquad (8)$$

and by nonstationary increments [1,6,7] we mean that

$$x(t + T) - x(t) \neq x(T). \qquad (9)$$

Again, such equations are to be understood as 'equality in distribution'. The implications of this distinction for data analysis, and for understanding Hurst exponents, are central. When (8) holds, then given the density of 'positions' $f_1(x,t)$, we also know the density f of increments $f(\Delta x)$,

$$f(x(t+T) - x(t)) = f_1(x,T), \qquad (10)$$

making it easy to *construct* what one means[1] by 'equality in distribution'. When the increments are nonstationary then it is impossible to construct a t-independent one-point density f of increments. The reason for this is that in the latter case a description of the increments inherently requires a two-point density, $f_2(x(t),t;x(t+T),t+T))$. So while equations such as (8) and (9) must always be understood as holding "in distribution", in the case of equation (9) there is no way to construct a t-indenpendent increment distribution. This is particularly true for Markov processes where $H \neq 1/2$.

By the efficient market hypothesis (EMH), we mean that the market is impossible to beat, that there are no correlations (no systematically repeated price/returns patterns) that can be exploited for profit. Such a market is necessarily uncorrelated noise, albeit not simply uncorrelated Gaussian noise. Because real markets are very hard to beat, models that generate no autocorrelations in increments are a good zeroth order approximation to real markets [7]. In such models, the autocorrelations in increments $\Delta x(t,T)=x(t+T)-x(t)$ vanish

$$\langle (x(t_1) - x(t_1 - T_1))(x(t_2 + T_2) - x(t_2)) \rangle = 0, \quad (11)$$

if there is no time interval overlap,

$$[t_1 - T_1, t_1] \cap [t_2, t_2 + T_2] = \Phi, \quad (12)$$

where $\Phi$ denotes the empty set on the line. This is a weaker condition than asserting that the increments are statistically independent.

---

[1] With a nonconsrtuctive definition or nonconstructive existence proof one is not quite sure what one is talking about.

Consider a stochastic process x(t) where the increments are uncorrelated. From this condition we easily obtain the autocorrelation function for positions (returns). Let t>s, then

$$\langle x(t)x(s) \rangle = \langle (x(t) - x(s))x(s) \rangle + \langle x^2(s) \rangle = \langle x^2(s) \rangle > 0, \quad (13)$$

since $x(s)-x(t_o)=x(s)$, so that $<x(s)x(t)>=<x^2(s)>=\sigma^2$ is simply the variance in x. All Markov processes will be seen by construction to generate this autocorrelation. If, in addition, scaling holds, then (13) yields

$$\langle x(t)x(s) \rangle = \min(s^{2H}, t^{2H}) \langle x^2(1) \rangle. \quad (14)$$

We next make a very important point. Combining

$$\langle (x(t+T) - x(t))^2 \rangle = +\langle x^2(t+T) \rangle + \langle x^2(t) \rangle - 2\langle x(t+T)x(t) \rangle$$
(15)

with (13), we get

$$\langle (x(t+T) - x(t))^2 \rangle = \langle x^2(t+T) \rangle - \langle x^2(t) \rangle \qquad (16)$$

which depends on both t and T, excepting the rare case where $\langle x^2(t) \rangle$ is linear in t. ***Uncorrelated increments are generally nonstationary.*** E.g., if scaling holds, then (16) becomes

$$\langle (x(t+T) - x(t))^2 \rangle = \langle x^2(1) \rangle ((t+T)^{2H} - t^{2H}). \qquad (17)$$

Here, the increments *may* be stationary iff. H=1/2, otherwise they're nonstationary. This class includes Markov processes. Next, we describe the class of stochastic processes that includes fractional Brownian motion (fBm), stochastic processes with arbitrarily long time memory.

Consider the class of all stochastic processes with stationary increments. Here, we begin with

$$-2\langle x(t+T)x(t) \rangle = \langle (x(t+T) - x(t))^2 \rangle - \langle x^2(t+T) \rangle - \langle x^2(t) \rangle, \qquad (18)$$

then using (8) on the right hand side of (18) we obtain

$$-2\langle x(t+T)x(t) \rangle = \langle x^2(T) \rangle - \langle x^2(t+T) \rangle - \langle x^2(t) \rangle \qquad (19)$$

which differs from (13). For increments with nonoverlapping time intervals, the simplest autocorrelation function is

$$2\langle (x(t) - x(t-T))(x(t+T) - x(t)) \rangle = \langle (x(t+T) - x(t-T))^2 \rangle - \langle (x(t) - x(t-T))^2 \rangle - \langle (x(t+T) - x(t))^2 \rangle$$
$$= \langle x^2(2T) \rangle - 2\langle x^2(T) \rangle$$
(20)

which generally does not vanish. ***Stationary increments are typically strongly correlated.*** E.g., if scaling (1) holds then we obtain the prediction of infinitely long time autocorrelations

$$\langle (x(t)-x(t-T))(x(t+T)-x(t)) \rangle = \langle x^2(T) \rangle (2^{2H-1}-1). \quad (21)$$

characteristic of fBm. This autocorrelation vanishes iff. H=1/2, otherwise the autocorrelations are strong for all time scales T. Such fluctuations violate the EMH, especially if H cannot be approximated as H≈1/2, and therefore could at best occur as higher order effects in finance markets.

Summarizing, the Hurst exponent H, taken alone, tells us nothing whatsoever about the autocorrelations, tells us nothing about the underlying dynamics. In the next two sections we will sharpen the distinction between scaling Markov processes and fBm where H≠1/2.

## 4. Markov Processes

We define Markov processes using densities and then show that the definition yields null autocorrelations for increments over nonoverlapping time intervals, showing that nonstationary Markov processes typically generate nonstationary increments.

A Markov process is a stochastic process without memory [10,11,12]: the conditional probability density for $x(t_n)$ in a time series $\{x(t_1), x(t_2), \ldots x(t_n)\}$ depends only on $x(t_{n-1})$, and so is independent of all earlier trajectory history $x_1, \ldots, x_{n-2}$. For finance markets, one best studies logarithmic returns [4,6] $x(t)=\ln p(t)/p_c$, where p is the price of a particular financial instrument. For a Markov process, the 2-point

probability density $f_2(x(t_1),t_1; x(t_2),t_2)$ is enough. We can then write

$$f_2(x(t),t;x(t+T),t+T) = g(x(t+T),t+T;x(t),t)f_1(x(t),t) \quad (22a)$$

where $f_1$ is a 1-point density of initial conditions and g is the transition density, or Green function. If we integrate over the variable at the earlier time in $f_2$, then follows that

$$f_1(x, t) = \int_{-\infty}^{\infty} dy\, g(x, t + T; y, t) f_1(y, t) \quad (22b)$$

A necessary condition for a Markov process is

$$g(x, t; x_o, t_o) = \int_{-\infty}^{\infty} dx'\, g(x, t; x', t') g(x', t'; x_o, t_o) \quad (23)$$

if $t_o<t'<t$. Next, drift-free motion, we show how the Markov property guarantees uncorrelated increments (11) over nonoverlapping time intervals.

First, note that if the increments are nonstationary, then even if we know the Green function g we don't know the corresponding density for the *increments*. To prove that a Markov process guarantess uncorrelated increments, we can formulate the problem as follows. We can prove lack of autocorrelations $<(x(t_1)-x(t_1-T))(x(t_2+T)-x(t_2))>$ for nonoverlapping time intervals $t_1<t_2$ with $T>0$, if we can show that the autocorrelations $<x(t)x(t+T)>$ reduce to the second moment at the shorter time (13). That is, with $T>0$ in

$$\langle x(t)x(t+T)\rangle = \int\int dx\,dy\,xy\,g(y,t+T;x,t)f_1(x,t), \quad (24)$$

we must show that $<x(t)x(t+T)>=<x^2(t)>$ if $T>0$.

We see that this is true for a Martingale (see Durrett [12] for Martingales), because the conditional average of x(t+T) starting from a point x(t) is then

$$\int dy\, y\, g(y, t+T; x, t) = x,  \qquad (25)$$

which yields

$$\langle x(t)x(t+T)\rangle = \int\int dx\, x^2 f_1(x,t) = \langle x^2(t)\rangle. \qquad (26)$$

for T>0. So if we work with the Martingale variable z(t)=x(t)-∫R(t)dt instead of x(t), then lack of autocorrelations of increments is proven. The stochastic differential equation for a drift free Markov process describbes a martingale.

The drift and diffusion coefficients are defined by

$$R(x,T,t) = \frac{\langle x(t+T) - x(t)\rangle}{T} = \frac{1}{T}\int (x-x_o)g(x,t+T;x_o,t)dx_o \qquad (27)$$

and

$$D(x,t) = \frac{\langle (x(t+T) - x(t))^2\rangle}{T} = \frac{1}{T}\int (x-x_o)^2 g(x,t+T;x_o,t)dx_o \qquad (28)$$

as T vanishes. However, one never knows the Green function in advance. Rather, one typically writes down a model diffusion coefficient D(x,t) and then calculates the density f(x,t). Diffusion coefficients can be inferred from Markov empirical data from a combination of the histograms (form of f as a function of x) and the variance (giving the time dependence in f).

*We conclude that Markov processes that scale with H≠1/2 will generate nonstationary, uncorrelated increments.*

## 5. Scaling Markov processes

We review scaling Markov processes [6] in order to contrast uncorrelated Gaussian Markov processes with fBm, which is a strongly correlated Gaussian process.

Here, we restrict to the density $f_1(x,t)=g(x,t;0,0)$ because the full Green function $g(x,t;x_o,t_o)$ does not scale [6].

A Markov process is generated locally by the stochastic diffferential equation (sde) [6]

$$dx = R(x,t) + \sqrt{D(x,t)}dB(t) \quad (29)$$

where B(t) is the Wiener process [12] and R a D are ordinary functions (not functionals) of the random variable x and time t. A Wiener process is an uncorrelated Gaussian process scaling with H=1/2, so that the increments are stationary and (from Ito's theorem [4,12]) $dB^2=dt=<dB^2>$.

The variance can be calculated from (29) as

$$\sigma^2 = \int_0^t ds \int_{-\infty}^{\infty} dx f_1(x,s)D(x,s), \quad (30)$$

so that scaling of the density and the variance imply that the diffusion coefficient scales as well [6]:

$$D(x,t) = t^{2H-1}D(u), u = x/t^H. \quad (31)$$

Hurst exponent scaling for Markov processes is possible with R(t) independent of x, but not with arbitrary drift R(x,t). One can derive a scaling requirement for a general drift R(x,t), but such scaling is generally not satisfied. In the remainder of the paper we'll assume a time dependent drift R(t) that has been removed, so that by x(t) we mean x(t) -∫R(t)dt. This means that we work with the drift free Fokker-Planck pde

$$\frac{\partial f_1}{\partial t} = \frac{1}{2}\frac{\partial^2}{\partial x^2}(Df_1). \qquad (32)$$

Scaling solutions for Markov processes are easily calculated [6,13,14]. With

$$f_1(x,t) = t^{-H}F(u); u = x/t^H \qquad (33)$$

and

$$D(x,t) = t^{2H-1}D(u), u = x/t^H \qquad (34)$$

the Fokker-Planck pde (32) yields

$$2H(uF(u))' + (D(u)F(u))'' = 0 \qquad (35)$$

which we integrate to obtain

$$F(u) = \frac{C}{D(u)}e^{-2H\int u du/D(u)} \qquad (36)$$

*For H≠1/2 all of these processes generate nonstationary increments.*

In particular, the choice $D(u)=D=$constant yields the Gaussian returns model $F(u)=(H/D\pi)^{1/2}\exp(-Hu^2/D)= (1/2\pi\langle x^2(1)\rangle)^{1/2}\exp(-u^2/2\langle x^2(1)\rangle)$ with $0<H<1$. This is second main point of this section, and is all that we need for this paper: *Gaussian Markov processs with H≠1/2 generate nonstationary increments*. But there are Gaussian processes that are not Markovian.

## 6. Fractional Brownian motion

We construct what is difficult (or impossible) to find written down explicitly in the literature, namely, the 1- and 2-point Gaussian densities of fBm. The two-point density, where the long-time autocorrelations are built in, does not scale with Hurst exponent H.

We can obtain scaling (1) from integrals of the type [2]

$$x_H(T) = \int_{-\infty}^{T} k(T,s)dB(s) \qquad (37)$$

if the kernel scales, $k(t,s)=t^{H-1/2}k(1,s/t)$, *and* if the lower limit of integration is, as shown, at $s=-\infty$. Long-time correlations for increments over nonoverlapping time intervals (21) follow if the increments of (37) are stationary. This can only be checked for a specific kernel. Mandelbrot and van Ness [2] have provided us with a scaling model with stationary increments,

$$x_H(t) = \int_{-\infty}^{0}[(t-s)^{H-1/2}-(-s)^{H-1/2}]dB(s) + \int_{0}^{t}[(t-s)^{H-1/2}dB(s), \quad (38)$$

a model of fBm with $\sigma^2(t)=\langle x^2(1)\rangle t^{2H}$. Deleting the first integral in (38) would yield a scaling process with nonstationary increments. We can also use the shorthand notation of the Ito product [4,15], $x(t)=k\bullet\Delta B$ for (38), where $t_o=-\infty$.

Here's where confusion may arise: if one calculates the 'one-point density' (which in reality is in this case a propagator from a *single* initial condition, $x(-\infty)=0$ to a present position $x(t)$) using

$$f_1(x,t) = \langle\delta(x - k\bullet\Delta B)\rangle = \frac{1}{2\pi}\int_{-\infty}^{\infty} e^{ipx}\langle e^{-ipk\bullet\Delta B}\rangle dp \qquad (39)$$

then one obtains a scaling Gaussian $f(x,t)=t^{-H}F(u)$, $F(u)=(1/2\pi\langle x^2(1)\rangle)^{1/2}\exp(-u^2/2\langle x^2(1)\rangle)$ with $u=x/t^H$, which is identical with the one point density of a scaling Gaussian Markov process. However, if one instead asssumes an initial time $t_o>-\infty$, $x(t_o)=0$, in (39), then one obtains a one point density that does not scale with H: it is necessary that $x(-\infty)=0$ in (38), otherwise the increments are not stationary. The two-point transition density cannot be used to construct a time evolution operator, the time evolution is described instead via the hierarchy of memory-dependent transition densities which reduce to a 2-point transition density iff. a process is Markovian.

So a Markov process cannot be distinguished from fBm on the basis of histograms (one-point densities) alone, it's necessary to ask if the increments are stationary or nonstationary when $H\neq 1/2$, and that is a question requiring two-point densities (for $H=1/2$ (38) is Markovian, is the Wiener process).

We can construct the two-point density that defines fBm. One needs a Gaussian process where scaling (1) holds, but with stationary increments [1]. Any two-point Gaussian density [11] is given by

$$f_2(x,t) = \frac{1}{2\pi \det B} e^{-x^+ B^{-1} x} \qquad (40)$$

where

$$B_{kl} = \langle x_k x_l \rangle. \qquad (41)$$

defines the autocorrelation matrix. Without specifying the autocorrelations (41), one cannot say whether the process x(t) is Markovian or not. The autocorrelation

$$\langle x(s)x(t) \rangle = \frac{\langle x^2(1) \rangle}{2} (|s|^{2H} + |t|^{2H} - |s-t|^{2H}) \qquad (42)$$

enforces stationary increments, where scaling with H is also asssumed in agreement with (38), and therefore will enforce the long time autocorrelations of fBm in the increments. The resulting two-point density of fBm can be written as[2]

$$f_2(x(s),s;x(t),t) = \frac{1}{2\pi \sigma_1 \sigma_2 (1-\rho^2)^{1/2}} e^{-(x^2(s)/\sigma_1^2 + x^2(t)/\sigma_2^2 - 2\rho x(s)xs(t)/\sigma_1\sigma_2)/2(1-\rho^2)}$$
(43)

where $\sigma_1\sigma_2\rho = <x(s)x(t)>$, $\sigma_1^2 = <x^2(1)>(abs(s))^{2H}$ and $\sigma_2^2 = <x^2(1)>(abs(t))^{2H}$ and

---

[2] This corrects misstatements about fBm in [4,6].

$$\rho = (|s|^{2H} + |t|^{2H} - |t-s|^{2H}) / 2|st|^H$$

If we integrate (43) over the earlier variable x(s), taking s<t, then we obtain the one point density $f_1$,

$$f_1(x,t,-\infty) = \frac{t^{-H}}{\sqrt{2\pi\langle x^2(1)\rangle}} e^{-x^2/2\langle x^2(1)\rangle t^{2H}}, \qquad (44)$$

which scales with H, is identical with the density for a Markov process, *but* where the initial condition must be understood to be x(-∞)=0. For a Markov process, in contrast, we can generally take x(0)=0.

Using (43) to construct the two point transition density, $p_2(y,s;x,t)=f_2(x,t;y,s)/f_1(x,t)$, one can show that fBm is not a Martingale: we obtain the conditional average <y(s)>=∫dyyp$_2$(y,s;x,t)=C(s,t)x, where C≠1 unless H=1/2. In fact, C is either positive or negative according to whether H is above or below ½ in value.

The implication of this paper for the analysis of time series should be clear: the two central questions are those of nonstationary vs. stationary increments, and correlated vs. uncorrelated increments. Scaling makes modelling easier but can't be counted on to exist in empirical data. All of this is illustrated in our recent finance data analysis [7]. We emphasize that (i) a Hurst exponent H, taken alone, tells us nothing about the dynamics, and (even worse) (b), a one-point density, taken alone, tells us little or nothing about the dynamics. It's absolutely necessary to study the autocorrelations of increments in order to obtain any idea what sort of dynamics are generated by financial (or any other) data.

## 7. Linear vs. Nonlinear Diffusion Revisited

Here is a clear description of Borland's scaling version [6,16] of Tsallis dynamics. Consider scaling Markov processes defined by (36), i.e., by linear diffusion. Ask for the class of diffusive scaling processes [6] $f(x,t)=t^{-H}F(u)$, $D(x,t)=t^{2H-1}D(u)$, where f is a power of D, $f(x,t)=D(x,t)^{1/(1-q)}$. The choice of the exponent in the form $1/(1-q)$ has no special significance other than it agrees with the notation of [16]. It follows that $(2H-1)/(1-q)=-H$, or $H=1/(3-q)$. Much more generally, we know that F(u) is a power of D(u) when the diffusion coefficient is quadratic in the scaling variable u [6], so that F is student t like: with

$$D(u) = d(\varepsilon)(1+\varepsilon u^2) \quad (45)$$

we obtain the general class of scaling student t like densities f(x,t), whereby

$$F(u) = C(1+\varepsilon u^2)^{-1-H/\varepsilon d(\varepsilon)}. \quad (46)$$

$H=1/(3-q)$ is only a special case. In general, H and ε are independent parameters. In the Borland-Tsallis model the Hurst exponent fixes both ε and the fat tail exponent μ. With $H=1/(3-q)$ we therefore obtain exactly the scaling density derived by Borland, who instead of the above derivation self-consistently solved the superficially nonlinear looking pde

$$\frac{\partial f}{\partial t} - \frac{1}{2}\frac{\partial^2}{\partial x^2}(Df) = \frac{\partial f}{\partial t} - \frac{1}{2}\frac{\partial^2}{\partial x^2}(f^{2-q}) = 0 \quad (47)$$

by assuming (45) and $f(x,t)=D(x,t)^{1/(1-q)}$ with $H=1/(3-q)$.

Clearly, there is no evidence for any specific 'nonlinear' behavior here, but the reader may ask: Is it possible that the density f(x,t) derived from linear diffusion (48,50) with $H=1/(3-q)$ can also represent a truly nonlinear solution derived from

$$\frac{\partial f}{\partial t} - \frac{1}{2}\frac{\partial^2}{\partial x^2}(f^{2-q}) = 0 \quad ? \quad (48)$$

Yes, the same density may represent entirely different stochastic processes, as we have shown above in the context of fBm, because a one point density tells us nothing about the underlying dynamics of a time series x(t). One point is that even truly nonlinear solutions of (51) will tell us nothing about memory in the form of increment autocorrelations. A more difficult point is that there is no way to identify a class of time series {x(t)} to which (51), without our linear interpretation, applies.

To answer questions about memory in the form of increment autocorrelations one would need to derive a two point or higher order density from a more general theory, a theory yielding (51) as the exact one point pde. Such theories may or may not exist, and if they do they may be may well be nonunique. The main point here is that one would have to derive correlations from a pde other than (48). No such calculation has been produced in the literature. *To date, all published results on Tsallis dynamics can be described as simple linear diffusion, no evidence for any specific nonlinear behavior has ever been produced*. It is, therefore, completely empty to argue that a student t like density represents the solution of a nonlinear system or describes 'nonlinear feedback'. Every stochastic differential equation with drift and diffusion coefficients not linear in x is a nonlinear sde, while a truly nonlinear pde has no underlying sde.

A nonlinear diffusion pde (48) for a one point density f(x,t) tells us nothing about the dynamics of any possible underlying time series x(t). A Langevin eqn. cannot be used to define the class of time series because nonlinear diffusion admits no Green function, whereas every Langevin eqn. with ordinary functions as drift and diffusion coefficients ($D=f^{1-q}$ is an ordinary function of (x,t), not a functional) describes a Markov process [6], and therefore generates a Green function (the transition probability density). The point is that we have no way to know what, if any, class of time series a more specific nonlinear system with the one point diffusion pde (51) might describe.

Finally, (49) yields the range of fat tail exponents $2<\mu<\infty$. For $\mu>3$ the variance is finte and scales with hurst exponent H [6], whereas for $2<\mu<3$ we obtain the Levy range of exponents and infinite variance but from a diffusive model. Again, the data indicate diffusive models with finite variance [7], ruling out Levy processes.

## Acknowledgement

KEB is supported by the NSF through grants #DMR-0406323 and #DMR-0427938, and the AFRL. GHG is supported by the NSF through grant #DMS-0607345 and by TLCC. JMC thanks Eduardo Roman for persistent email that focused him this line of thought, is very grateful to Enrico Scalas for persistently critical and very motivating email, and to Harry

Thomas for helpful papers on stochastic systems with memory [15]. We're grateful to a referee for useful remarks.